\documentclass[aps,prx,a4paper,nofootinbib,superscriptaddress,numbers,longbibliography,showpacs,showkeys,floatfix,svgnames,preprint,reprint,nobalancelastpage]{revtex4-2}

\usepackage{amsfonts,amssymb,amsmath}
\usepackage{hyperref,xcolor,graphicx}
\usepackage[separate-uncertainty=true]{siunitx}

\DeclareSIUnit\gauss{G}

\usepackage[english]{babel}
\usepackage[utf8]{inputenc}
\usepackage[T1]{fontenc}

\usepackage{xspace}

\usepackage{physics}

\newcommand{\figref}[2]{\hyperref[#1]{\ref{#1}{(#2)}}}
\addto\captionsenglish{}
\addto\captionsenglish{}

\newcommand\Tstrut{\rule{0pt}{2.6ex}}         \newcommand\Bstrut{\rule[-0.9ex]{0pt}{0pt}}   

\def\theTitle{%
Ramsey imaging of optical traps
}

\hypersetup{
 pdftitle = {\theTitle},
 pdfauthor = {Gautam Ramola, Richard Winkelmann, Karthik Chandrashekara, Wolfgang Alt, Xu Peng, Dieter Meschede, Andrea Alberti},
	colorlinks=true,linkcolor=blue,citecolor=blue,filecolor=blue,urlcolor=blue,pdfpagemode=UseNone,pdfstartview={XYZ null null 1.00}
}

\renewcommand\textemdash{\leavevmode\unskip\kern0.8pt---\kern1pt\ignorespaces}

\makeatletter
\let\old@Section@Cmd=\section
\def\end@of@sec@tit{.\leavevmode\unskip\kern0.8pt\rule[0.19\baselineskip]{8pt}{0.4pt}\kern1pt}
\def\mysection{%
	\def\reserved@stsec##1{\@startsection{section}{1}{\parindent}{\z@}{0em}{\normalfont\normalsize\itshape}*[##1]{##1\end@of@sec@tit}}%
	\@ifstar{%
		\reserved@stsec%
	}{%
		\reserved@stsec%
	}%
}
\let\section=\mysection
\makeatother

\begin{document}

\title{\theTitle}

\author{Gautam Ramola}
 \author{Richard Winkelmann}
\author{Karthik Chandrashekara}
\author{Wolfgang Alt}
\affiliation{Institut für Angewandte Physik, Universität Bonn, Wegelerstraße 8, 53115 Bonn, Germany}
\author{Xu Peng (\raisebox{-0.7mm}{\includegraphics[width=6.5mm]{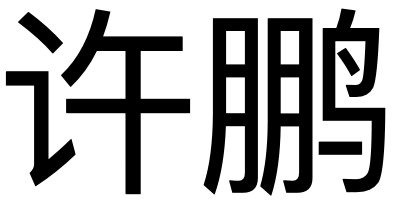}})}
\affiliation{Institut für Angewandte Physik, Universität Bonn, Wegelerstraße 8, 53115 Bonn, Germany}
\affiliation{State Key Laboratory of Magnetic Resonance and Atomic and Molecular Physics, and Wuhan National Laboratory for Optoelectronics, Wuhan Institute of
Physics and Mathematics, Chinese Academy of Sciences, Wuhan 430071, China}
\author{Dieter Meschede}
\affiliation{Institut für Angewandte Physik, Universität Bonn, Wegelerstraße 8, 53115 Bonn, Germany}
\author{Andrea Alberti}
 \email{alberti@iap.uni-bonn.de}
\affiliation{Institut für Angewandte Physik, Universität Bonn, Wegelerstraße 8, 53115 Bonn, Germany}

\date{\today}             
\def\work{work}

\begin{abstract}
Mapping the potential landscape with high spatial resolution is crucial for quantum technologies based on ultracold atoms.
Yet, imaging optical dipole traps is challenging because purely optical methods, commonly used to profile laser beams in free space, are not applicable in vacuum.
In this \work, we demonstrate precise \emph{in-situ} imaging of optical dipole traps by probing a hyperfine transition with Ramsey interferometry.
Thereby, we obtain an absolute map of the potential landscape with micrometer resolution and shot-noise-limited spectral precision.
The idea of the technique is to control the polarization ellipticity of the trap laser beam to induce a differential light shift proportional to the trap potential.
By studying the response to polarization ellipticity, we uncover a small but significant nonlinearity in addition to a dominant linear behavior, which is explained by the geometric distribution of the atomic ensemble.
Our technique for imaging of optical traps can find wide application in quantum technologies based on ultracold atoms, as it applies to multiple atomic species and is not limited to a particular wavelength or trap geometry.
\end{abstract}

                              \maketitle

\section{Introduction}
\label{sec:introduction}

Controlling the potential landscape of optical traps is essential for atom-based quantum technologies such as optical lattice clocks \cite{Marti:2018}, trapped atom interferometers \cite{Poli:2011,Xu:2019b}, measurement-based quantum computing \cite{Briegel:2009,Yang:2020d}, analogue \cite{Bloch:2012a,Browaeys:2020} and discrete \cite{Sajid:2019} quantum simulators.
In these experiments, atoms are trapped in the dipole potential created by focused laser beams.
Precise knowledge of their intensity profile is indispensable to realize homogeneous potential landscapes with optical lattices~\cite{Heinz:2021} and arrays of micro traps~\cite{Barredo:2016,Endres:2016}, and it can also be used to avoid inhomogeneous line broadening \cite{Marti:2018}.
Laser beam profiling techniques, however, are not applicable in vacuum.
Instead, one directly probes the atomic ensemble itself in order to obtain \emph{in-situ} information about the trap potential, e.g., by measuring motional resonances spectroscopically or by direct imaging of the trapped atom density.
These indirect methods estimate the local potential based on some additional assumption (e.g., geometry of the beams, harmonic approximation, thermalization of the atomic ensemble),
and are not suitable to isolate the contribution of individual beams in setups involving multiple crossed beams, as is the case of optical lattices.

In contrast, techniques for imaging optical traps aim to directly map the local potential as experienced by the atoms.
For alkali-earth atoms, imaging of an optical trap was demonstrated by measuring the differential light shift induced by the trap laser field upon an ultranarrow optical transition \cite{Marti:2018}.
For species lacking narrow optical transitions, such as alkali atoms, related previous work showed that the scalar \cite{Bertoldi:2010} and vector \cite{Vengalattore:2007} differential light shifts induced by a dipole trap close to resonance can be used to image the local potential, but did not provide an absolute measurement thereof.

In this \work, we report on a technique for imaging optical traps with very high precision by probing a hyperfine transition of alkali atoms with position-resolved Ramsey phase tracking.
The Ramsey signal measures the differential light shift caused by a small, controllable polarization ellipticity of the trap laser beam, which is directly related to the trap potential as experienced by the atoms.
By integrating the signal over few repetitions, we attain a spectral resolution of about two orders of magnitude below the Fourier limit $\nu_F = 1/t$, where $t$ is the interrogation time.
Such a high resolution allows us to uncover a nonlinear response of the atomic ensemble to polarization ellipticity, which yields a significant systematic correction to the imaged optical trap.
As a result, we obtain an absolute map of the potential landscape at the micrometer scale with an accuracy of the order of \SI{e-2} of the maximum potential depth.
To exemplify the versatility of the Ramsey imaging technique, we map the dipole potential produced by individual laser beams propagating in diverse directions and with various wavelengths: in-between the D doublet (\SI{866}{\nano\meter}) and far detuned from it (\SI{1064}{\nano\meter}).

\section{Conceptual scheme}
\label{sec:concept}

\begin{figure}[tb]
\includegraphics[width=\columnwidth]{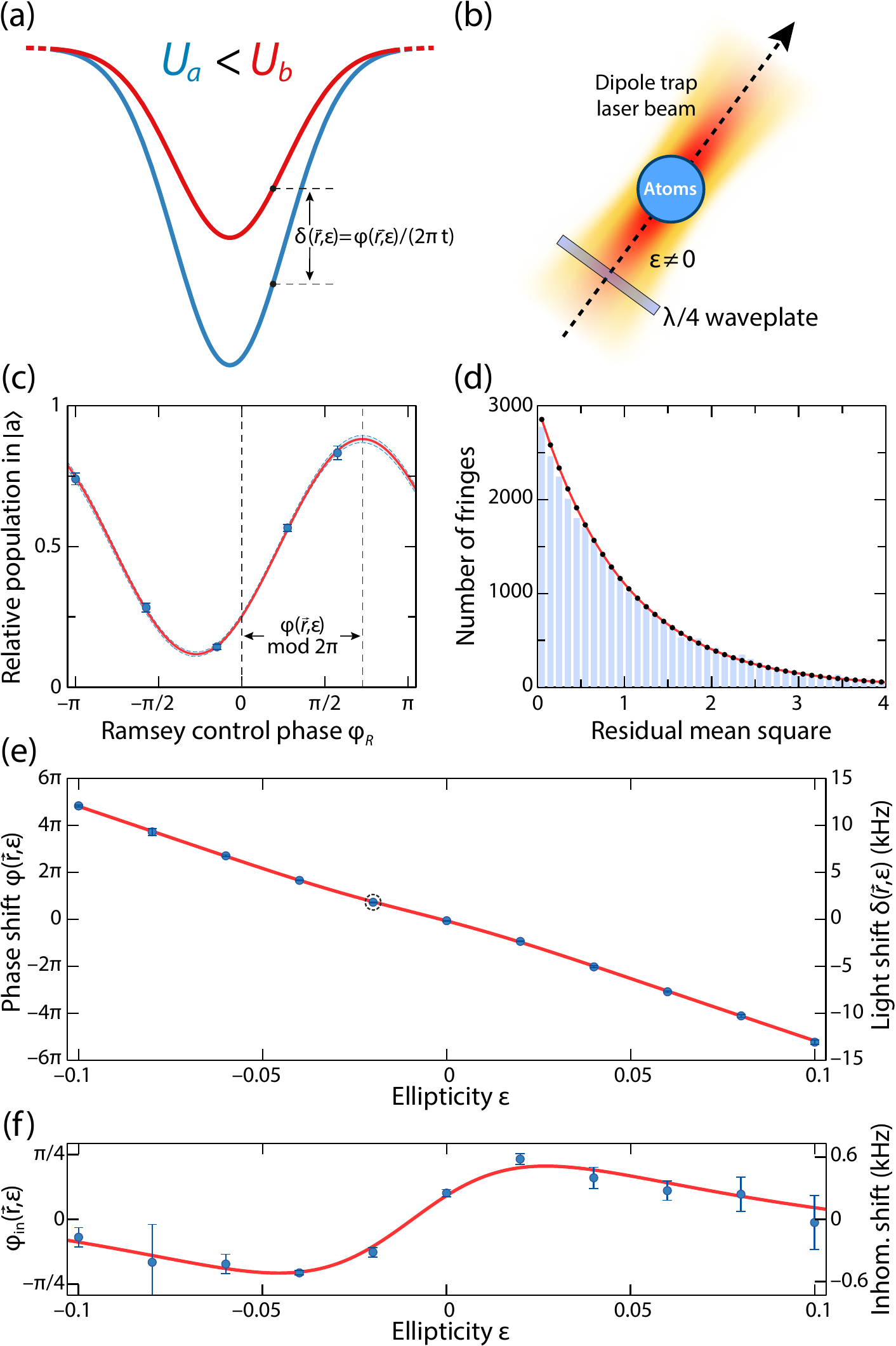}
\caption{\label{fig:concept}
Imaging of optical potentials by position-resolved Ramsey phase tracking.
(a) Atoms in states $\ket{a}$ and $\ket{b}$ experience different trap potentials because of differential light shift; see Eq.~(\ref{deltaProp}).
(b) The trap laser beam to be imaged is made elliptically polarized ($\epsilon\neq 0$) before impinging onto the atomic ensemble.
(c) A typical Ramsey interference fringe as a function of the control phase $\varphi_R$.
(d) Histogram of the residual mean squares derived from fitting the Ramsey fringes recorded for different $\vec{r}$ and $\epsilon$, in excellent agreement with the expected reduced $\chi^2$ distribution (solid line). 
(e) Ramsey phase as a function of $\epsilon$ measured for a given $\vec{r}$ [see marked pixel in Fig.~\figref{fig:beamImage}{c}] and a Ramsey duration $t=\SI{200}{\micro\second}$.
On the right y-axis, the values are given in frequency units, based on the linear relation in Eq.~(\ref{eq:linear_relation}).
The circled point is obtained from the Ramsey fringe in (c).
(f) The nonlinear contribution to the differential light shift is singled out by subtracting the linear contribution from the data points in (e).
Experimental data in (c-f) refer to the optical trap produced by the beam called H3 (see text), for which $\eta_s=\num{2.5e-3}$ and $\eta_v = \num{1.75}$.
}
\end{figure}

In an optical dipole trap, the differential light shift between two long-lived states $\ket{a}$ and $\ket{b}$
[Fig.~\figref{fig:concept}{a}] can be written as:
\begin{equation}
\delta(\vec{r},\epsilon) = (\eta_s+\eta_v\epsilon)\,U(\vec{r})/(2\pi\hbar),
\label{deltaProp}
\end{equation}
where $U(\vec{r})$ is the potential experienced by the atoms at position $\vec{r}$ for linear polarization,
$\eta_s$ and $\eta_v$ are two proportionality factors accounting for the scalar \cite{Kuhr:2005} and vector \cite{Le-Kien:2013} contribution \cite{lanthanide}, $\epsilon$ is the polarization ellipticity of the trap laser beam, and $\hbar$ is the reduced Planck constant.
The two factors $\eta_s$ and $\eta_v$ only depend on the laser beam's wavelength $\lambda$ and the atomic properties, whereas the polarization ellipticity is defined as
$\epsilon = (I_{R} - I_{L})/(I_{L} + I_{R})$, with $I_L$ and $I_R$ being the intensities of the left and right circular polarization components of the trap laser beam.
To obtain a map of the potential landscape $U(\vec{r})$, we exploit the linear dependence of the light shift $\delta(\vec{r},\epsilon)$ on the ellipticity $\epsilon$, which we can easily vary using a $\lambda/4$ waveplate [Fig.~\figref{fig:concept}{b}].
The light shift is measured precisely with Ramsey interferometry, by tracking the phase shift,
\begin{equation}
	\label{eq:linear_relation}
	\varphi(\vec{r},\epsilon) = 2\pi \, t \, \delta(\vec{r},\epsilon),
\end{equation}
of the Ramsey fringe [Fig.~\figref{fig:concept}{c}] for different values of $\epsilon$.

Alternatively, one can measure the shift $\delta(\vec{r},\epsilon)$ by directly recording a spectrum of the transition.
Ramsey interferometry, however, presents the advantage that the model function fitted to the data is known exactly in the form of a sinusoidal fringe.
By contrast, direct spectroscopy requires an accurate modeling of the line shape in order to determine the frequency shift with high precision.
To assess the quality of the Ramsey model function, we consider the distribution of the residual mean squares derived from fitting $>\num{e4}$ Ramsey fringes, probed at different locations $\vec{r}$ and for different values of $\epsilon$ [Fig.~\figref{fig:concept}{d}].
The comparison with the theoretical $\chi^2$ distribution shows an excellent agreement, validating the interpretation of the estimated fringe parameters as the most likely ones.
Assuming that the dominant noise source is atom shot noise, the uncertainty of the phase obtained from the fit can be expressed as
\begin{equation}
\label{eq:phaseError}
\varphi_\text{err}(\epsilon) = \frac{1}{[({1-\sqrt{1-C(\epsilon)^2})N}]^{1/2}},
\end{equation}
where $N$ is the number of atoms interrogated in the surrounding of $\vec{r}$, and $C(\epsilon)$ is the contrast of the Ramsey fringe, which is a decreasing function of $|\epsilon|$, as shown later.
For our measurements with $N\lesssim 500$, we find that the statistical fit uncertainty on $\varphi(\vec{r},\epsilon)$ is about a factor 2 greater than $\varphi_\text{err}(\epsilon)$.
The small excess noise could be explained by atom losses, photon shot noise and, to a lesser extent, by the read-out noise of the CCD camera.
The measured phase uncertainty translates through Eq.~(\ref{eq:linear_relation}) to a frequency uncertainty of about two orders of magnitude below $\nu_F$.

The measured Ramsey phase is found to be approximately proportional to the ellipticity $\epsilon$ [Fig.~\figref{fig:concept}{e}].
By subtracting the leading linear contribution, a small nonlinearity becomes evident [Fig.~\figref{fig:concept}{f}], which will be discussed in detail later.
If we ignore for now the nonlinear contribution, it is straightforward to obtain by linear extrapolation the phase $\varphi(\vec{r},\pm1)$ corresponding to a pure right or left circular polarization, $\epsilon=\pm 1$, and the phase $\varphi(\vec{r},0)$ corresponding to linear polarization.
The local trap potential can therefore be expressed as
\begin{equation}
	\label{eq:potential}
	U(\vec{r}) = \pm \frac{\varphi(\vec{r},\pm)-\varphi(\vec{r},0)}{S},
\end{equation} 
with the factor $S=\eta_v\,t/\hbar$ playing the role of the sensitivity factor of the Ramsey imaging technique.
We remark that the scalar contribution to the differential light shift, $\eta_s U(\vec{r})/(2\pi\hbar)$,
has no effect on the reconstructed potential landscape because it does not depend on $\epsilon$.

\section{Alkali atoms}
\label{sec:alkali_atoms}

We discuss the general case of an alkali atom, with $\ket{a} = \ket{F,m_F}$ and $\ket{b}=\ket{F',m'_F}$ being two hyperfine levels of the electronic ground state.
Here, $F$ and $F'$ denote the quantum numbers of the total angular momentum of the atom, whereas $m_F$ and $m_F'$ represent the corresponding magnetic quantum numbers, with the quantization axis aligned in the direction of the laser beam whose potential we aim to image. 
Accounting for the vector polarizability \cite{Grimm:2000}, we obtain the following expression:
\begin{equation}
	\label{eq:eta}
		\eta_v = (g'_F m'_F-g_F m_F)\frac{\nu_2-\nu_1}{2(\nu-\nu_1)+\nu-\nu_2},
\end{equation}%
where $g_F$ and $g'_F$ are the Landé factors of the states $\ket{a}$ and $\ket{b}$, $\nu_1$ and $\nu_2$ are the resonance frequencies of the $\mathrm{D}_1$ and $\mathrm{D}_2$ lines, and $\nu = c/\lambda$ is the frequency of the laser beam, with $c$ being the speed of light.

Equation~(\ref{eq:eta}) shows that the sensitivity factor $S$ is nonzero for all transitions except for clock-type transitions (when $g'_Fm'_F= g_Fm_F$).
Moreover, $S$ increases with the fine structure splitting, $\nu_2 - \nu_1$, which is larger for heavier atoms (Table~\ref{table:eta}), and for wavelengths closer to one of the two D lines.
This behavior is caused by the fact that Ramsey imaging of optical traps works for alkali atoms by leveraging the spin-orbit interaction present in the p-orbitals.

\begin{table}[t]
\caption{\label{table:eta}%
Reference values of $\eta_v$ and $\eta_s$ for a far-detuned dipole trap at \SI{1064}{\nano\meter} for three representative alkali atoms. $\eta_v$ is calculated for doubly polarized and maximally stretched hyperfine states of the electronic ground state. $\eta_s$ is much smaller than $\eta_v$ because it originates from the hyperfine interaction in the electronic ground state.}
\begin{ruledtabular}
\begin{tabular}{lccc}
\textrm{}&\textrm{${}^{133}$Cs}& \textrm{${}^{87}$Rb}& \textrm{${}^{23}$Na} \Tstrut\Bstrut \\ 
\hline
$\eta_v$ & \num{-0.16} &\num{-0.04}& \num{-1e-3} \Tstrut\\  
$\eta_s$ & \num{1.5e-4} &\num{6.8e-5} &\num{7.8e-6} \Bstrut\\ 
\end{tabular}
\end{ruledtabular}
\end{table}

\section{Experimental setup}
\label{sec:experiment}

We demonstrate Ramsey imaging of optical traps with ${}^{133}$Cs atoms,  probing the hyperfine transition between $\ket{a}=\ket{F=3,m_F=3}$ and $\ket{b}=\ket{F=4,m_F=4}$.
We map the potential landscape of four optical traps, which are produced by laser beams propagating in different directions and having different wavelengths.
Three of the beams (labeled H1, H2, H3, with $\lambda = \SI{866}{\nano\meter}$) propagate in a horizontal plane, which is perpendicular to the imaging axis, whereas the fourth beam (labeled V, with $\lambda=\SI{1064}{\nano\meter}$) propagates along the vertical direction, which coincides with the imaging axis, and is retro-reflected to form an optical standing wave.
 The four laser beams are overlapped to create a three-dimensional optical lattice, as detailed in Ref.~\cite{Groh16}.

For each beam of which we intend to map the potential, we make its polarization slightly elliptical by inserting a $\lambda/4$ waveplate in the beam path, as illustrated in Fig.~\figref{fig:concept}{b}.
The precise value of $\epsilon$ can be measured with standard ellipsometry methods, e.g., using a rotating polarizer.
As an alternative, in the case of beams H1 and H3, we use a digital polarization synthesizer \cite{RobensPolSynth} to control their ellipticity.

For a fixed value of $\epsilon$, the experimental sequence begins by cooling a cloud of several thousand of atoms in a magneto-optical trap.
Subsequently, the atoms are transferred into the foregoing three-dimensional optical lattice. The loading procedure is performed in such a way that the atoms are distributed over a relatively large region ($\SI{60}{\micro\meter}\times\SI{60}{\micro\meter}$),
covering \SI{90}{\percent} of the laser beams’ cross section.
The atoms trapped in the optical lattice are then further cooled in all three dimensions to $T\approx \SI{1}{\micro\kelvin}$ by resolved-sideband cooling \cite{Belmechri:2013}, and simultaneously pumped optically into state $\ket{b}$.
Importantly, the bias magnetic field of $\approx \SI{3}{\gauss}$, which defines the quantization axis, is rotated adiabatically in the direction of the laser beam to be imaged, by controlling the current flowing through three pairs of compensation coils.

For the Ramsey interferometer, we abruptly turn off all trap laser beams except for the one relevant to the measurement, 
and subsequently use microwave radiation to apply two short $\pi/2$ pulses of $\SI{1}{\micro\second}$ duration each, separated by a fixed interrogation time $t=\SI{200}{\micro\second}$.
The pulse frequency $\nu_\text{hfs}$ is chosen to be approximately resonant with the hyperfine transition between $\ket{a}$ and $\ket{b}$ for the atoms occupying the center of the trap.
Because of their short duration, the pulses are spectrally broad enough to allow the entire ensemble of atoms to be addressed, including the atoms in the outer regions of the trap where the differential light shift is much weaker.
To avoid systematic phase shifts that could arise in the outer regions of the trap due to slightly off-resonance pulses, we also record for beam H3 a reference Ramsey phase map for a vanishing time $t$, which is then subtracted.
For our experimental parameters, this correction is found to be nonsignificant.
We also remark that instead of microwave pulses, one could alternatively use optical Raman pulses, since any systematic phase shift caused by spatial intensity inhomogeneity of the intensity of the Raman laser beams can likewise be subtracted.

For the state-selective detection, we remove atoms in state $\ket{b}$ with an optical push-out pulse and then acquire a fluorescence image \cite{Alberti:2016} of the remaining atoms in state $\ket{a}$ through a high-numerical-aperture objective lens \cite{Robens:2017a}, which is well suited to resolve the atoms' positions with high precision.
We note that in the absence of an optical lattice, one can employ other imaging techniques (e.g., absorption imaging).
The fraction of remaining atoms exhibits a typical Ramsey fringe [Fig.~\figref{fig:concept}{c}],
\begin{equation}
P_{a}(\vec{r},\epsilon,\varphi_R) = \frac{1}{2}+\frac{C_0}{2}\cos[\varphi'(\vec{r},\epsilon) - \varphi_R]\,,
\label{eq:populationspindown}
\end{equation}
as we scan $\varphi_R$, the relative phase between the two $\pi/2$ pulses.
Here, $C_0$ is the fringe contrast, and $\varphi'(\vec{r},\epsilon)$ is a position-dependent phase obtained by tracking the shift of the Ramsey fringe.
Considering Eq.~(\ref{eq:linear_relation}), we thus find that the measured phase shift $\varphi'(\vec{r},\epsilon)$ directly yields the differential light shift: $\delta(\vec{r},\epsilon) = \varphi'(\vec{r},\epsilon)/(2\pi\,t) + \nu_\text{hfs}-\nu_{\text{hfs},0}(\vec{r})$, where $\nu_{\text{hfs},0}(\vec{r})$ is the frequency of the hyperfine transition in the absence of light fields.
The bare frequency $\nu_{\text{hfs},0}(\vec{r})$ generally varies with $\vec{r}$ due to residual magnetic field gradients, but importantly does not depend on $\epsilon$.
We thus simply use $\varphi'(\vec{r},\epsilon )$ in lieu of $\varphi(\vec{r},\epsilon)$ in Eq.~(\ref{eq:potential}) in order to obtain $U(\vec{r})$, without requiring any independent measurement of $\nu_{\text{hfs},0}(\vec{r})$.

\section{Experimental results}

The Ramsey signal is analyzed in a position-resolved manner by subdividing the field of view in small pixels.
For our analysis, we choose the pixel size $\Delta_P \approx \SI{3.6}{\micro\meter}$, corresponding to 10 pixels of the electron-multiplying CCD camera.
Owing to the nonvanishing temperature $T$, the atoms move during the Ramsey time $t$ on average by $\approx\SI{1}{\micro\meter}$, which is less than the pixel size $\Delta_P$.
We note that for a desired spatial resolution of the potential map, the thermal motion of atoms constrains the maximum allowed Ramsey interrogation time $t$, which in turn sets the Fourier limit $\nu_F$, and therefore eventually limits the spectral resolution.
There is thus a trade-off between the resolvable pixel size $\Delta_P$ and Fourier resolution $\nu_F$, where the product of these two quantities is proportional to $\sqrt{k_B T/m}$, with $m$ being the mass of the atoms and $k_B$ the Boltzmann constant.
We also note that any nonzero ellipticity causes a differential force on the states $\ket{a}$ and $\ket{b}$, leading to a displacement between the two wave packets during the interrogation time.
Such a displacement may cause a reduction of the contrast $C_0$, and induce a systematic phase shift.
We find, however, that the effect of the differential force is negligible in our case.

\begin{figure}[t]
\includegraphics[width=\columnwidth]{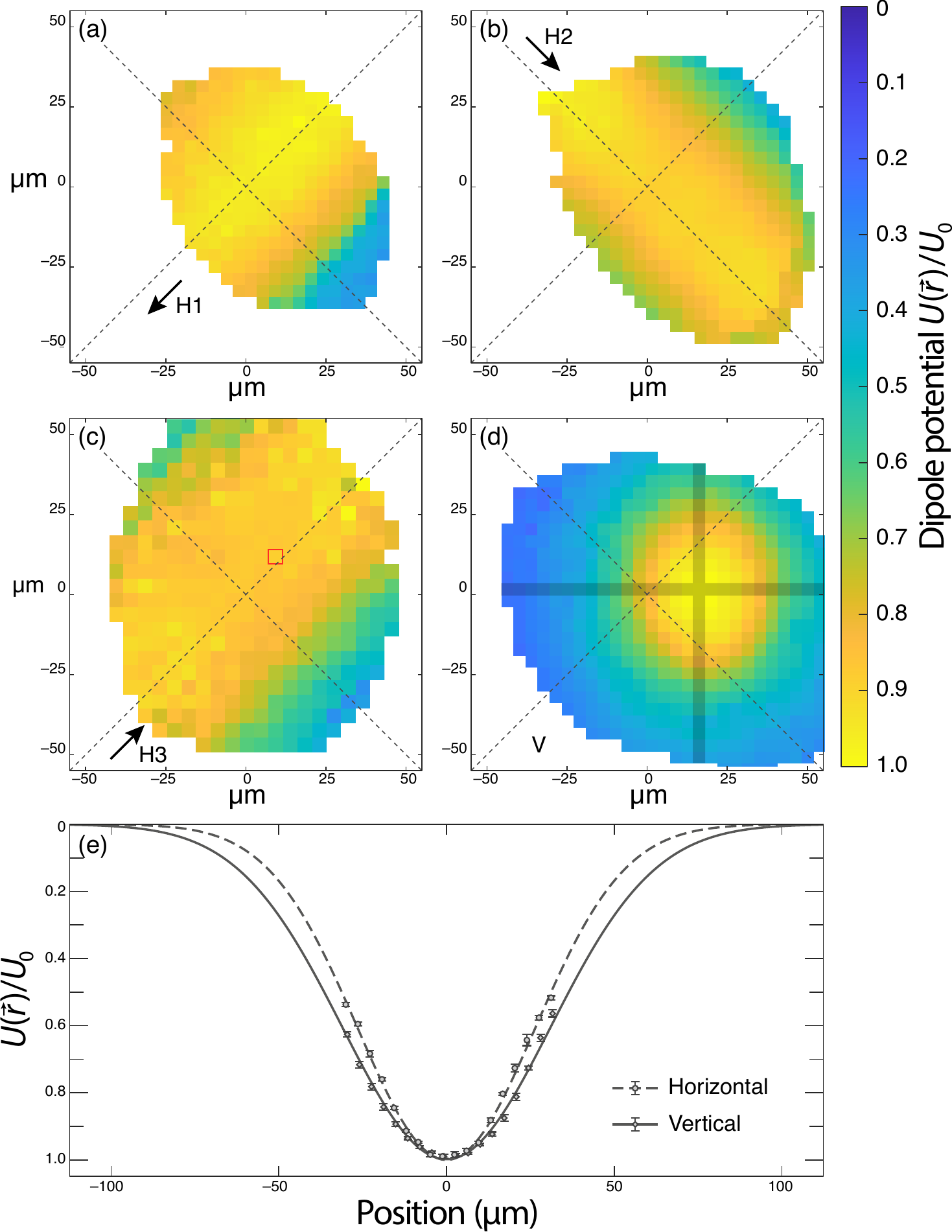}
\caption{\label{fig:beamImage}
Ramsey imaging of four optical traps. Reconstructed potential produced by (a-c) three laser beams with $\lambda=\SI{866}{\nano\meter}$ propagating in a common horizontal plane, and (d) a standing wave with $\lambda=\SI{1064}{\nano\meter}$ oriented along the line of sight.
The potential is expressed relative to the minimum of the potential, $U_0$.
The dashed lines represent common reference axes.
Shown in white are the outer regions where few or no atoms are loaded.
The arrows in (a-c) indicate the propagation direction of the running-wave laser beams,
whereas the marked pixel in (c) corresponds to the data shown in Fig.~\ref{fig:concept}.
(e) The data points correspond to two orthogonal cuts as highlighted in (d), 
whereas the lines are the two Gaussian fitting functions.
The slightly different trap waists are likely caused by a small astigmatism of beam V.
}
\end{figure}

Figure~\ref{fig:beamImage} shows the potential map for each of the four optical traps, obtained using position-resolved Ramsey phase tracking.
Observing the positions of the four reconstructed trap potentials, it is immediately noticeable that the V beam is off-centered by about $\SI{16}{\micro\meter}$ from the intersection point of the three horizontal beams.
A more careful analysis of individual transverse cuts of the mapped potentials shows that the trap waist and the potential depth are determined with an uncertainty of $<\SI{1}{\percent}$, whereas the trap transverse position with an uncertainty of a few hundred nanometers.
Two orthogonal cuts along the principal axes of beam V are shown in Fig.~\figref{fig:beamImage}{e} as a representative example, revealing two slightly different trap waists.
Such precise geometric information about the position and waist of optical traps is crucial for the sensitive alignment of multiple laser beams in vacuum, and would be hard, if not practically unfeasible to accomplish using purely optical methods or with indirect \emph{in-situ} measurements of the atomic ensemble.

\begin{figure}[t]
\includegraphics[width=\columnwidth]{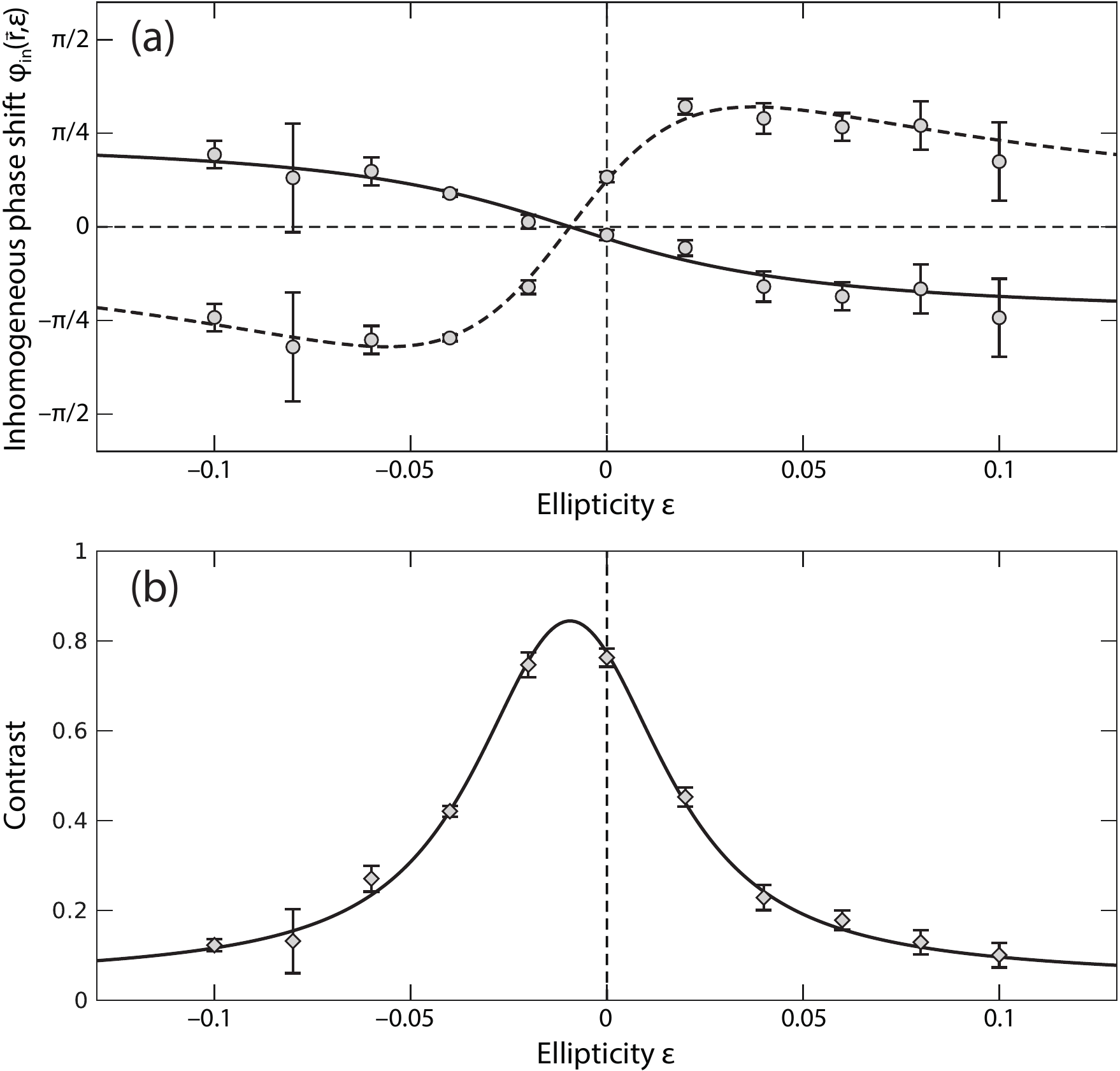}
\caption{\label{fig:inhomogenous}
Nonlinear response of Ramsey signal to polarization ellipticity $\epsilon$, measured for a single representative pixel, marked in Fig.~\figref{fig:beamImage}{c}. (a) Same data as in Fig.~\figref{fig:concept}{f} showing the nonlinear phase correction.
For a close comparison, we display the two nonlinear terms appearing in Eq.~(\ref{eq:phaseterm}) separately by a dashed line (first term) and a solid line (second term).
The data points are obtained by subtracting all terms of the fitting model, except for the one of interest.
(b) The Ramsey fringe contrast as a function $\epsilon$ for the same points in (a), showing a very good agreement with the model in Eq.~(\ref{eq:contrastterm}).
An additional ellipticity offset of $\approx 0.01$ is used to fit the data in (a) and (b), which likely originates from a systemic error in the ellipticity measurement.
}
\end{figure}

\section{Nonlinear response model}

While the linear model in Eq.~(\ref{deltaProp}) captures the essence of the Ramsey imaging technique, this model alone is not sufficient to explain the small but significant nonlinear corrections observed as a function of $\epsilon$ in Fig.~\figref{fig:concept}{f}.
We attribute this nonlinear behavior to the distribution of atoms along the direction collinear with the line of sight.

In this direction, the position of atoms cannot be resolved directly by the imaging system.
For any given pixel, the detected Ramsey fringe is thus the integrated result of many atoms, each experiencing a slightly different trap laser intensity and, consequently, an inhomogeneous differential light shift.
This inhomogenous differential light shift has a two-fold effect on the Ramsey signal: it causes a contrast reduction of the Ramsey fringe and a nonlinear shift, $\varphi_\text{in}$, of the Ramsey phase as a function of $\epsilon$.

To quantitatively model these two effects, we consider a single pixel, and assume a one-dimensional Gaussian distribution for the positions of the atoms along the the line of sight.
The resulting Ramsey fringe is thus obtained by averaging over the ensemble:
\begin{multline}
	\label{eq:ramsey_average}
  \int_{-\infty}^{\infty}\hspace{-3pt}e^{i\varphi(\vec{r},\epsilon)}\, p(z-z_0,\sigma) \, \mathrm{d}z = \\ =C(\epsilon) \, e^{i (\eta_s+\eta_v\epsilon)\hspace{0.3pt}U\hspace{0.3pt}t/h}e^{i\varphi_\text{in}(\epsilon)},
\end{multline} 
where $z$ is the coordinate along the line of sight, $p(z-z_0,\sigma)$ is a Gaussian function  centered at $z_0$ with width $\sigma$, and $U$ is the maximum trap depth along the line of sight, for the pixel considered.
To compute the integral in Eq.~(\ref{eq:ramsey_average}) analytically, we assume that the trap potential along the $z$-direction can be approximated by a harmonic oscillator of angular frequency $\omega$.
The computation shows that the phase correction and the contrast can be modeled as:
\begin{subequations}
\label{eq:nonlinear}
\begin{flalign}
\label{eq:phaseterm}
\varphi_\text{in}(\epsilon) &= \frac{\xi(\epsilon)}{1+\xi(\epsilon)^2} \left(\frac{z_0}{\sqrt{2}\sigma}\right)^2\hspace{-2pt}-\arctan\left[\xi(\epsilon)\right]\hspace{-1pt}, \\
\label{eq:contrastterm}
C(\epsilon) &= \frac{C_0}{\left[1+\xi(\epsilon)^2\right]^{1/4}}\exp\hspace{-2pt}\left[-\frac{\xi(\epsilon)^2}{1+\xi(\epsilon)^2}\hspace{-3pt}\left(\frac{z_0}{\sqrt{2}\sigma}\right)^{\hspace{-3pt}2}\right]\hspace{-1pt},
\end{flalign}
\end{subequations}
where $\xi(\epsilon)=(\eta_s+\eta_v \epsilon)\,m\omega^2\sigma^2/(\hbar/t)$ is a dimensionless quantity linear in the ellipticity $\epsilon$.
We fit this model simultaneously to both the nonlinear phase contribution [Fig.~\figref{fig:inhomogenous}{a}] and the contrast [Fig.~\figref{fig:inhomogenous}{b}], which are measured as a function of $\epsilon$ for a single representative pixel of beam H3.
The measured data are in remarkable agreement with the fitting model, from which we obtain that the spread $\sigma$ is between \SIrange{10}{15}{\micro\meter}, depending on the pixel position.

Studying the fitting model in Eqs.~(\ref{eq:phaseterm}) and (\ref{eq:contrastterm}), one recognizes that the nonlinear behavior of the phase correction manifests itself only when the fringe contrast $C$ has dropped significantly.
It is therefore important to vary $\epsilon$ over a sufficiently wide range in order to discriminate the nonlinear correction from the leading linear contribution of Eq.~(\ref{eq:linear_relation}).
As a result of the nonlinear model, we obtain an absolute map of the potential landscape, which is found to be about \SI{10}{\percent} deeper than that obtained using the linear model in Eq.~(\ref{eq:linear_relation}).

\section{Conclusions}
\label{sec:conclusions}
We have demonstrated a technique for the precise imaging of optical trap potentials by Ramsey interferometry.
For our demonstration, we have mapped the potential landscape of optical traps for Cs atoms, considering different geometries and wavelengths.
Ramsey imaging of optical traps can find application to other atomic species, such as alkali atoms and magnetic lanthanides \cite{Kao:2017,Becher:2018,Ravensbergen:2018}, and to a variety of optical trap geometries (e.g., optical lattices, flat traps, hollow traps, tailored potentials).

\acknowledgments{The authors would like to thank Carsten Robens for helpful discussions.
This research was supported by the SFB/TR 185 OSCAR of the German Research Foundation.
G.~R.\ acknowledges support from the Bonn-Cologne Graduate School of Physics and Astronomy.
}

\cleardoublepage

\bibliographystyle{apsrev4-2}
\bibliography{beamProfilingBib}

\providecommand{\noopsort}[1]{}\providecommand{\singleletter}[1]{#1}%
\begin{thebibliography}{25}%
\makeatletter
\providecommand \@ifxundefined [1]{%
 \@ifx{#1\undefined}
}%
\providecommand \@ifnum [1]{%
 \ifnum #1\expandafter \@firstoftwo
 \else \expandafter \@secondoftwo
 \fi
}%
\providecommand \@ifx [1]{%
 \ifx #1\expandafter \@firstoftwo
 \else \expandafter \@secondoftwo
 \fi
}%
\providecommand \natexlab [1]{#1}%
\providecommand \enquote  [1]{``#1''}%
\providecommand \bibnamefont  [1]{#1}%
\providecommand \bibfnamefont [1]{#1}%
\providecommand \citenamefont [1]{#1}%
\providecommand \href@noop [0]{\@secondoftwo}%
\providecommand \href [0]{\begingroup \@sanitize@url \@href}%
\providecommand \@href[1]{\@@startlink{#1}\@@href}%
\providecommand \@@href[1]{\endgroup#1\@@endlink}%
\providecommand \@sanitize@url [0]{\catcode `\\12\catcode `\$12\catcode
  `\&12\catcode `\#12\catcode `\^12\catcode `\_12\catcode `\%12\relax}%
\providecommand \@@startlink[1]{}%
\providecommand \@@endlink[0]{}%
\providecommand \url  [0]{\begingroup\@sanitize@url \@url }%
\providecommand \@url [1]{\endgroup\@href {#1}{\urlprefix }}%
\providecommand \urlprefix  [0]{URL }%
\providecommand \Eprint [0]{\href }%
\providecommand \doibase [0]{https://doi.org/}%
\providecommand \selectlanguage [0]{\@gobble}%
\providecommand \bibinfo  [0]{\@secondoftwo}%
\providecommand \bibfield  [0]{\@secondoftwo}%
\providecommand \translation [1]{[#1]}%
\providecommand \BibitemOpen [0]{}%
\providecommand \bibitemStop [0]{}%
\providecommand \bibitemNoStop [0]{.\EOS\space}%
\providecommand \EOS [0]{\spacefactor3000\relax}%
\providecommand \BibitemShut  [1]{\csname bibitem#1\endcsname}%
\let\auto@bib@innerbib\@empty
\bibitem [{\citenamefont {Marti}\ \emph {et~al.}(2018)\citenamefont {Marti},
  \citenamefont {Hutson}, \citenamefont {Goban}, \citenamefont {Campbell},
  \citenamefont {Poli},\ and\ \citenamefont {Ye}}]{Marti:2018}%
  \BibitemOpen
  \bibfield  {author} {\bibinfo {author} {\bibfnamefont {G.~E.}\ \bibnamefont
  {Marti}}, \bibinfo {author} {\bibfnamefont {R.~B.}\ \bibnamefont {Hutson}},
  \bibinfo {author} {\bibfnamefont {A.}~\bibnamefont {Goban}}, \bibinfo
  {author} {\bibfnamefont {S.~L.}\ \bibnamefont {Campbell}}, \bibinfo {author}
  {\bibfnamefont {N.}~\bibnamefont {Poli}},\ and\ \bibinfo {author}
  {\bibfnamefont {J.}~\bibnamefont {Ye}},\ }\bibfield  {title} {\enquote
  {\bibinfo {title} {{Imaging Optical Frequencies with \SI{100}{\micro\hertz}
  Precision and \SI{1.1}{\micro\meter} Resolution}},}\ }\href
  {https://doi.org/10.1103/PhysRevLett.120.103201} {\bibfield  {journal}
  {\bibinfo  {journal} {Phys. Rev. Lett.}\ }\textbf {\bibinfo {volume} {120}},\
  \bibinfo {pages} {103201} (\bibinfo {year} {2018})}\BibitemShut {NoStop}%
\bibitem [{\citenamefont {Poli}\ \emph {et~al.}(2011)\citenamefont {Poli},
  \citenamefont {Wang}, \citenamefont {Tarallo}, \citenamefont {Alberti},
  \citenamefont {Prevedelli},\ and\ \citenamefont {Tino}}]{Poli:2011}%
  \BibitemOpen
  \bibfield  {author} {\bibinfo {author} {\bibfnamefont {N.}~\bibnamefont
  {Poli}}, \bibinfo {author} {\bibfnamefont {F.~Y.}\ \bibnamefont {Wang}},
  \bibinfo {author} {\bibfnamefont {M.~G.}\ \bibnamefont {Tarallo}}, \bibinfo
  {author} {\bibfnamefont {A.}~\bibnamefont {Alberti}}, \bibinfo {author}
  {\bibfnamefont {M.}~\bibnamefont {Prevedelli}},\ and\ \bibinfo {author}
  {\bibfnamefont {G.~M.}\ \bibnamefont {Tino}},\ }\bibfield  {title} {\enquote
  {\bibinfo {title} {{Precision Measurement of Gravity with Cold Atoms in an
  Optical Lattice and Comparison with a Classical Gravimeter}},}\ }\href
  {https://doi.org/10.1103/PhysRevLett.106.038501} {\bibfield  {journal}
  {\bibinfo  {journal} {Phys. Rev. Lett.}\ }\textbf {\bibinfo {volume} {106}},\
  \bibinfo {pages} {038501} (\bibinfo {year} {2011})}\BibitemShut {NoStop}%
\bibitem [{\citenamefont {Xu}\ \emph {et~al.}(2019)\citenamefont {Xu},
  \citenamefont {Jaffe}, \citenamefont {Panda}, \citenamefont {Kristensen},
  \citenamefont {Clark},\ and\ \citenamefont {M{\"u}ller}}]{Xu:2019b}%
  \BibitemOpen
  \bibfield  {author} {\bibinfo {author} {\bibfnamefont {V.}~\bibnamefont
  {Xu}}, \bibinfo {author} {\bibfnamefont {M.}~\bibnamefont {Jaffe}}, \bibinfo
  {author} {\bibfnamefont {C.~D.}\ \bibnamefont {Panda}}, \bibinfo {author}
  {\bibfnamefont {S.~L.}\ \bibnamefont {Kristensen}}, \bibinfo {author}
  {\bibfnamefont {L.~W.}\ \bibnamefont {Clark}},\ and\ \bibinfo {author}
  {\bibfnamefont {H.}~\bibnamefont {M{\"u}ller}},\ }\bibfield  {title}
  {\enquote {\bibinfo {title} {{Probing gravity by holding atoms for 20
  seconds}},}\ }\href {https://doi.org/10.1126/science.aay6428} {\bibfield
  {journal} {\bibinfo  {journal} {Science}\ }\textbf {\bibinfo {volume}
  {366}},\ \bibinfo {pages} {745} (\bibinfo {year} {2019})}\BibitemShut
  {NoStop}%
\bibitem [{\citenamefont {Briegel}\ \emph {et~al.}(2009)\citenamefont
  {Briegel}, \citenamefont {Browne}, \citenamefont {D{\"u}r}, \citenamefont
  {Raussendorf},\ and\ \citenamefont {den Nest}}]{Briegel:2009}%
  \BibitemOpen
  \bibfield  {author} {\bibinfo {author} {\bibfnamefont {H.~J.}\ \bibnamefont
  {Briegel}}, \bibinfo {author} {\bibfnamefont {D.~E.}\ \bibnamefont {Browne}},
  \bibinfo {author} {\bibfnamefont {W.}~\bibnamefont {D{\"u}r}}, \bibinfo
  {author} {\bibfnamefont {R.}~\bibnamefont {Raussendorf}},\ and\ \bibinfo
  {author} {\bibfnamefont {M.~V.}\ \bibnamefont {den Nest}},\ }\bibfield
  {title} {\enquote {\bibinfo {title} {{Measurement-based quantum
  computation}},}\ }\href {https://doi.org/10.1038/nphys1157} {\bibfield
  {journal} {\bibinfo  {journal} {Nat. Phys.}\ }\textbf {\bibinfo {volume}
  {5}},\ \bibinfo {pages} {19} (\bibinfo {year} {2009})}\BibitemShut {NoStop}%
\bibitem [{\citenamefont {Yang}\ \emph {et~al.}(2020)\citenamefont {Yang},
  \citenamefont {Sun}, \citenamefont {Huang}, \citenamefont {Wang},
  \citenamefont {Deng}, \citenamefont {Dai}, \citenamefont {Yuan},\ and\
  \citenamefont {Pan}}]{Yang:2020d}%
  \BibitemOpen
  \bibfield  {author} {\bibinfo {author} {\bibfnamefont {B.}~\bibnamefont
  {Yang}}, \bibinfo {author} {\bibfnamefont {H.}~\bibnamefont {Sun}}, \bibinfo
  {author} {\bibfnamefont {C.-J.}\ \bibnamefont {Huang}}, \bibinfo {author}
  {\bibfnamefont {H.-Y.}\ \bibnamefont {Wang}}, \bibinfo {author}
  {\bibfnamefont {Y.}~\bibnamefont {Deng}}, \bibinfo {author} {\bibfnamefont
  {H.-N.}\ \bibnamefont {Dai}}, \bibinfo {author} {\bibfnamefont {Z.-S.}\
  \bibnamefont {Yuan}},\ and\ \bibinfo {author} {\bibfnamefont {J.-W.}\
  \bibnamefont {Pan}},\ }\bibfield  {title} {\enquote {\bibinfo {title}
  {{Cooling and entangling ultracold atoms in optical lattices}},}\ }\href
  {https://doi.org/10.1126/science.aaz6801} {\bibfield  {journal} {\bibinfo
  {journal} {Science}\ }\textbf {\bibinfo {volume} {369}},\ \bibinfo {pages}
  {550} (\bibinfo {year} {2020})}\BibitemShut {NoStop}%
\bibitem [{\citenamefont {Bloch}\ \emph {et~al.}(2012)\citenamefont {Bloch},
  \citenamefont {Dalibard},\ and\ \citenamefont
  {Nascimb{\`e}ne}}]{Bloch:2012a}%
  \BibitemOpen
  \bibfield  {author} {\bibinfo {author} {\bibfnamefont {I.}~\bibnamefont
  {Bloch}}, \bibinfo {author} {\bibfnamefont {J.}~\bibnamefont {Dalibard}},\
  and\ \bibinfo {author} {\bibfnamefont {S.}~\bibnamefont {Nascimb{\`e}ne}},\
  }\bibfield  {title} {\enquote {\bibinfo {title} {{Quantum simulations with
  ultracold quantum gases}},}\ }\href {https://doi.org/10.1038/nphys2259}
  {\bibfield  {journal} {\bibinfo  {journal} {Nat. Phys.}\ }\textbf {\bibinfo
  {volume} {8}},\ \bibinfo {pages} {267} (\bibinfo {year} {2012})}\BibitemShut
  {NoStop}%
\bibitem [{\citenamefont {Browaeys}\ and\ \citenamefont
  {Lahaye}(2020)}]{Browaeys:2020}%
  \BibitemOpen
  \bibfield  {author} {\bibinfo {author} {\bibfnamefont {A.}~\bibnamefont
  {Browaeys}}\ and\ \bibinfo {author} {\bibfnamefont {T.}~\bibnamefont
  {Lahaye}},\ }\bibfield  {title} {\enquote {\bibinfo {title} {{Many-body
  physics with individually controlled Rydberg atoms}},}\ }\href
  {https://doi.org/10.1038/s41567-019-0733-z} {\bibfield  {journal} {\bibinfo
  {journal} {Nat. Phys.}\ }\textbf {\bibinfo {volume} {16}},\ \bibinfo {pages}
  {132} (\bibinfo {year} {2020})},\ \Eprint {https://arxiv.org/abs/2002.07413}
  {2002.07413} \BibitemShut {NoStop}%
\bibitem [{\citenamefont {Sajid}\ \emph {et~al.}(2019)\citenamefont {Sajid},
  \citenamefont {Asb{\'o}th}, \citenamefont {Meschede}, \citenamefont
  {Werner},\ and\ \citenamefont {Alberti}}]{Sajid:2019}%
  \BibitemOpen
  \bibfield  {author} {\bibinfo {author} {\bibfnamefont {M.}~\bibnamefont
  {Sajid}}, \bibinfo {author} {\bibfnamefont {J.~K.}\ \bibnamefont
  {Asb{\'o}th}}, \bibinfo {author} {\bibfnamefont {D.}~\bibnamefont
  {Meschede}}, \bibinfo {author} {\bibfnamefont {R.~F.}\ \bibnamefont
  {Werner}},\ and\ \bibinfo {author} {\bibfnamefont {A.}~\bibnamefont
  {Alberti}},\ }\bibfield  {title} {\enquote {\bibinfo {title} {{Creating
  anomalous Floquet Chern insulators with magnetic quantum walks}},}\ }\href
  {https://doi.org/10.1103/PhysRevB.99.214303} {\bibfield  {journal} {\bibinfo
  {journal} {Phys. Rev. B}\ }\textbf {\bibinfo {volume} {99}},\ \bibinfo
  {pages} {214303} (\bibinfo {year} {2019})}\BibitemShut {NoStop}%
\bibitem [{\citenamefont {Heinz}\ \emph {et~al.}(2021)\citenamefont {Heinz},
  \citenamefont {Trautmann}, \citenamefont {{\v S}anti{\'c}}, \citenamefont
  {Jihyun~Park}, \citenamefont {Bloch},\ and\ \citenamefont
  {Blatt}}]{Heinz:2021}%
  \BibitemOpen
  \bibfield  {author} {\bibinfo {author} {\bibfnamefont {A.}~\bibnamefont
  {Heinz}}, \bibinfo {author} {\bibfnamefont {J.}~\bibnamefont {Trautmann}},
  \bibinfo {author} {\bibfnamefont {N.}~\bibnamefont {{\v S}anti{\'c}}},
  \bibinfo {author} {\bibfnamefont {A.}~\bibnamefont {Jihyun~Park}}, \bibinfo
  {author} {\bibfnamefont {I.}~\bibnamefont {Bloch}},\ and\ \bibinfo {author}
  {\bibfnamefont {S.}~\bibnamefont {Blatt}},\ }\bibfield  {title} {\enquote
  {\bibinfo {title} {{Crossed optical cavities with large mode diameters}},}\
  }\href {https://doi.org/10.1364/OL.414076} {\bibfield  {journal} {\bibinfo
  {journal} {Opt. Lett.}\ }\textbf {\bibinfo {volume} {46}},\ \bibinfo {pages}
  {250} (\bibinfo {year} {2021})}\BibitemShut {NoStop}%
\bibitem [{\citenamefont {Barredo}\ \emph {et~al.}(2016)\citenamefont
  {Barredo}, \citenamefont {de~L{\'e}s{\'e}leuc}, \citenamefont {Lienhard},
  \citenamefont {Lahaye},\ and\ \citenamefont {Browaeys}}]{Barredo:2016}%
  \BibitemOpen
  \bibfield  {author} {\bibinfo {author} {\bibfnamefont {D.}~\bibnamefont
  {Barredo}}, \bibinfo {author} {\bibfnamefont {S.}~\bibnamefont
  {de~L{\'e}s{\'e}leuc}}, \bibinfo {author} {\bibfnamefont {V.}~\bibnamefont
  {Lienhard}}, \bibinfo {author} {\bibfnamefont {T.}~\bibnamefont {Lahaye}},\
  and\ \bibinfo {author} {\bibfnamefont {A.}~\bibnamefont {Browaeys}},\
  }\bibfield  {title} {\enquote {\bibinfo {title} {{An atom-by-atom assembler
  of defect-free arbitrary two-dimensional atomic arrays}},}\ }\href
  {https://doi.org/10.1126/science.aah3778} {\bibfield  {journal} {\bibinfo
  {journal} {Science}\ }\textbf {\bibinfo {volume} {354}},\ \bibinfo {pages}
  {1021} (\bibinfo {year} {2016})}\BibitemShut {NoStop}%
\bibitem [{\citenamefont {Endres}\ \emph {et~al.}(2016)\citenamefont {Endres},
  \citenamefont {Bernien}, \citenamefont {Keesling}, \citenamefont {Levine},
  \citenamefont {Anschuetz}, \citenamefont {Krajenbrink}, \citenamefont
  {Senko}, \citenamefont {Vuletic}, \citenamefont {Greiner},\ and\
  \citenamefont {Lukin}}]{Endres:2016}%
  \BibitemOpen
  \bibfield  {author} {\bibinfo {author} {\bibfnamefont {M.}~\bibnamefont
  {Endres}}, \bibinfo {author} {\bibfnamefont {H.}~\bibnamefont {Bernien}},
  \bibinfo {author} {\bibfnamefont {A.}~\bibnamefont {Keesling}}, \bibinfo
  {author} {\bibfnamefont {H.}~\bibnamefont {Levine}}, \bibinfo {author}
  {\bibfnamefont {E.~R.}\ \bibnamefont {Anschuetz}}, \bibinfo {author}
  {\bibfnamefont {A.}~\bibnamefont {Krajenbrink}}, \bibinfo {author}
  {\bibfnamefont {C.}~\bibnamefont {Senko}}, \bibinfo {author} {\bibfnamefont
  {V.}~\bibnamefont {Vuletic}}, \bibinfo {author} {\bibfnamefont
  {M.}~\bibnamefont {Greiner}},\ and\ \bibinfo {author} {\bibfnamefont {M.~D.}\
  \bibnamefont {Lukin}},\ }\bibfield  {title} {\enquote {\bibinfo {title}
  {{Atom-by-atom assembly of defect-free one-dimensional cold atom arrays}},}\
  }\href {https://doi.org/10.1126/science.aah3752} {\bibfield  {journal}
  {\bibinfo  {journal} {Science}\ }\textbf {\bibinfo {volume} {354}},\ \bibinfo
  {pages} {1024} (\bibinfo {year} {2016})}\BibitemShut {NoStop}%
\bibitem [{\citenamefont {Bertoldi}\ \emph {et~al.}(2010)\citenamefont
  {Bertoldi}, \citenamefont {Bernon}, \citenamefont {Vanderbruggen},
  \citenamefont {Landragin},\ and\ \citenamefont {Bouyer}}]{Bertoldi:2010}%
  \BibitemOpen
  \bibfield  {author} {\bibinfo {author} {\bibfnamefont {A.}~\bibnamefont
  {Bertoldi}}, \bibinfo {author} {\bibfnamefont {S.}~\bibnamefont {Bernon}},
  \bibinfo {author} {\bibfnamefont {T.}~\bibnamefont {Vanderbruggen}}, \bibinfo
  {author} {\bibfnamefont {A.}~\bibnamefont {Landragin}},\ and\ \bibinfo
  {author} {\bibfnamefont {P.}~\bibnamefont {Bouyer}},\ }\bibfield  {title}
  {\enquote {\bibinfo {title} {{In situ characterization of an optical cavity
  using atomic light shift}},}\ }\href {https://doi.org/10.1364/OL.35.003769}
  {\bibfield  {journal} {\bibinfo  {journal} {Opt. Lett.}\ }\textbf {\bibinfo
  {volume} {35}},\ \bibinfo {pages} {3769} (\bibinfo {year}
  {2010})}\BibitemShut {NoStop}%
\bibitem [{\citenamefont {Vengalattore}\ \emph {et~al.}(2007)\citenamefont
  {Vengalattore}, \citenamefont {Higbie}, \citenamefont {Leslie}, \citenamefont
  {Guzman}, \citenamefont {Sadler},\ and\ \citenamefont
  {Stamper-Kurn}}]{Vengalattore:2007}%
  \BibitemOpen
  \bibfield  {author} {\bibinfo {author} {\bibfnamefont {M.}~\bibnamefont
  {Vengalattore}}, \bibinfo {author} {\bibfnamefont {J.~M.}\ \bibnamefont
  {Higbie}}, \bibinfo {author} {\bibfnamefont {S.~R.}\ \bibnamefont {Leslie}},
  \bibinfo {author} {\bibfnamefont {J.}~\bibnamefont {Guzman}}, \bibinfo
  {author} {\bibfnamefont {L.~E.}\ \bibnamefont {Sadler}},\ and\ \bibinfo
  {author} {\bibfnamefont {D.~M.}\ \bibnamefont {Stamper-Kurn}},\ }\bibfield
  {title} {\enquote {\bibinfo {title} {{High-Resolution Magnetometry with a
  Spinor Bose-Einstein Condensate}},}\ }\href
  {https://doi.org/10.1103/PhysRevLett.98.200801} {\bibfield  {journal}
  {\bibinfo  {journal} {Phys. Rev. Lett.}\ }\textbf {\bibinfo {volume} {98}},\
  \bibinfo {pages} {200801} (\bibinfo {year} {2007})}\BibitemShut {NoStop}%
\bibitem [{\citenamefont {Kuhr}\ \emph {et~al.}(2005)\citenamefont {Kuhr},
  \citenamefont {Alt}, \citenamefont {Schrader}, \citenamefont {Dotsenko},
  \citenamefont {Miroshnychenko}, \citenamefont {Rauschenbeutel},\ and\
  \citenamefont {Meschede}}]{Kuhr:2005}%
  \BibitemOpen
  \bibfield  {author} {\bibinfo {author} {\bibfnamefont {S.}~\bibnamefont
  {Kuhr}}, \bibinfo {author} {\bibfnamefont {W.}~\bibnamefont {Alt}}, \bibinfo
  {author} {\bibfnamefont {D.}~\bibnamefont {Schrader}}, \bibinfo {author}
  {\bibfnamefont {I.}~\bibnamefont {Dotsenko}}, \bibinfo {author}
  {\bibfnamefont {Y.}~\bibnamefont {Miroshnychenko}}, \bibinfo {author}
  {\bibfnamefont {A.}~\bibnamefont {Rauschenbeutel}},\ and\ \bibinfo {author}
  {\bibfnamefont {D.}~\bibnamefont {Meschede}},\ }\bibfield  {title} {\enquote
  {\bibinfo {title} {{Analysis of dephasing mechanisms in a standing-wave
  dipole trap}},}\ }\href {https://doi.org/10.1103/PhysRevA.72.023406}
  {\bibfield  {journal} {\bibinfo  {journal} {Phys. Rev. A}\ }\textbf {\bibinfo
  {volume} {72}},\ \bibinfo {pages} {023406} (\bibinfo {year}
  {2005})}\BibitemShut {NoStop}%
\bibitem [{\citenamefont {Le~Kien}\ \emph {et~al.}(2013)\citenamefont
  {Le~Kien}, \citenamefont {Schneeweiss},\ and\ \citenamefont
  {Rauschenbeutel}}]{Le-Kien:2013}%
  \BibitemOpen
  \bibfield  {author} {\bibinfo {author} {\bibfnamefont {F.}~\bibnamefont
  {Le~Kien}}, \bibinfo {author} {\bibfnamefont {P.}~\bibnamefont
  {Schneeweiss}},\ and\ \bibinfo {author} {\bibfnamefont {A.}~\bibnamefont
  {Rauschenbeutel}},\ }\bibfield  {title} {\enquote {\bibinfo {title}
  {{Dynamical polarizability of atoms in arbitrary light fields: general theory
  and application to cesium}},}\ }\href
  {https://doi.org/10.1140/epjd/e2013-30729-x} {\bibfield  {journal} {\bibinfo
  {journal} {Eur. Phys. J. D.}\ }\textbf {\bibinfo {volume} {67}},\ \bibinfo
  {pages} {92} (\bibinfo {year} {2013})}\BibitemShut {NoStop}%
\bibitem [{lan()}]{lanthanide}%
  \BibitemOpen
  \bibinfo {note} {If $\ket{a}$ and $\ket{b}$ have total electronic angular
  momentum $J \geq 1/2$ (e.g., ground state levels of magnetic lanthanides), an
  additional contribution to Eq.~(\ref{deltaProp}) originates from the tensor
  polarizability \cite{Le-Kien:2013}. This contribution, however, does not
  dependent on the polarization when the quantization axis is chosen collinear
  with the laser beam direction. Therefore, the tensor contribution can be
  effectively absorbed in $\eta_s$.}\BibitemShut {Stop}%
\bibitem [{\citenamefont {Grimm}\ \emph {et~al.}(2000)\citenamefont {Grimm},
  \citenamefont {Weidem{\"u}ller},\ and\ \citenamefont
  {Ovchinnikov}}]{Grimm:2000}%
  \BibitemOpen
  \bibfield  {author} {\bibinfo {author} {\bibfnamefont {R.}~\bibnamefont
  {Grimm}}, \bibinfo {author} {\bibfnamefont {M.}~\bibnamefont
  {Weidem{\"u}ller}},\ and\ \bibinfo {author} {\bibfnamefont {Y.~B.}\
  \bibnamefont {Ovchinnikov}},\ }\bibfield  {title} {\enquote {\bibinfo {title}
  {{Optical Dipole Traps for Neutral Atoms}},}\ }\href
  {https://doi.org/10.1016/S1049-250X(08)60186-X} {\bibfield  {journal}
  {\bibinfo  {journal} {Adv. At. Mol. Opt. Phy.}\ }\textbf {\bibinfo {volume}
  {42}},\ \bibinfo {pages} {95} (\bibinfo {year} {2000})}\BibitemShut {NoStop}%
\bibitem [{\citenamefont {Groh}\ \emph {et~al.}(2016)\citenamefont {Groh},
  \citenamefont {Brakhane}, \citenamefont {Alt}, \citenamefont {Meschede},
  \citenamefont {Asb\'oth},\ and\ \citenamefont {Alberti}}]{Groh16}%
  \BibitemOpen
  \bibfield  {author} {\bibinfo {author} {\bibfnamefont {T.}~\bibnamefont
  {Groh}}, \bibinfo {author} {\bibfnamefont {S.}~\bibnamefont {Brakhane}},
  \bibinfo {author} {\bibfnamefont {W.}~\bibnamefont {Alt}}, \bibinfo {author}
  {\bibfnamefont {D.}~\bibnamefont {Meschede}}, \bibinfo {author}
  {\bibfnamefont {J.~K.}\ \bibnamefont {Asb\'oth}},\ and\ \bibinfo {author}
  {\bibfnamefont {A.}~\bibnamefont {Alberti}},\ }\bibfield  {title} {\enquote
  {\bibinfo {title} {Robustness of topologically protected edge states in
  quantum walk experiments with neutral atoms},}\ }\href
  {https://doi.org/10.1103/PhysRevA.94.013620} {\bibfield  {journal} {\bibinfo
  {journal} {Phys. Rev. A}\ }\textbf {\bibinfo {volume} {94}},\ \bibinfo
  {pages} {013620} (\bibinfo {year} {2016})}\BibitemShut {NoStop}%
\bibitem [{\citenamefont {Robens}\ \emph {et~al.}(2018)\citenamefont {Robens},
  \citenamefont {Brakhane}, \citenamefont {Alt}, \citenamefont {Meschede},
  \citenamefont {Zopes},\ and\ \citenamefont {Alberti}}]{RobensPolSynth}%
  \BibitemOpen
  \bibfield  {author} {\bibinfo {author} {\bibfnamefont {C.}~\bibnamefont
  {Robens}}, \bibinfo {author} {\bibfnamefont {S.}~\bibnamefont {Brakhane}},
  \bibinfo {author} {\bibfnamefont {W.}~\bibnamefont {Alt}}, \bibinfo {author}
  {\bibfnamefont {D.}~\bibnamefont {Meschede}}, \bibinfo {author}
  {\bibfnamefont {J.}~\bibnamefont {Zopes}},\ and\ \bibinfo {author}
  {\bibfnamefont {A.}~\bibnamefont {Alberti}},\ }\bibfield  {title} {\enquote
  {\bibinfo {title} {Fast, high-precision optical polarization synthesizer for
  ultracold-atom experiments},}\ }\href
  {https://doi.org/10.1103/PhysRevApplied.9.034016} {\bibfield  {journal}
  {\bibinfo  {journal} {Phys. Rev. Applied}\ }\textbf {\bibinfo {volume} {9}},\
  \bibinfo {pages} {034016} (\bibinfo {year} {2018})}\BibitemShut {NoStop}%
\bibitem [{\citenamefont {Belmechri}\ \emph {et~al.}(2013)\citenamefont
  {Belmechri}, \citenamefont {F{\"o}rster}, \citenamefont {Alt}, \citenamefont
  {Widera}, \citenamefont {Meschede},\ and\ \citenamefont
  {Alberti}}]{Belmechri:2013}%
  \BibitemOpen
  \bibfield  {author} {\bibinfo {author} {\bibfnamefont {N.}~\bibnamefont
  {Belmechri}}, \bibinfo {author} {\bibfnamefont {L.}~\bibnamefont
  {F{\"o}rster}}, \bibinfo {author} {\bibfnamefont {W.}~\bibnamefont {Alt}},
  \bibinfo {author} {\bibfnamefont {A.}~\bibnamefont {Widera}}, \bibinfo
  {author} {\bibfnamefont {D.}~\bibnamefont {Meschede}},\ and\ \bibinfo
  {author} {\bibfnamefont {A.}~\bibnamefont {Alberti}},\ }\bibfield  {title}
  {\enquote {\bibinfo {title} {{Microwave control of atomic motional states in
  a spin-dependent optical lattice}},}\ }\href
  {https://doi.org/10.1088/0953-4075/46/10/104006} {\bibfield  {journal}
  {\bibinfo  {journal} {J. Phys. B: At. Mol. Phys.}\ }\textbf {\bibinfo
  {volume} {46}},\ \bibinfo {pages} {104006} (\bibinfo {year}
  {2013})}\BibitemShut {NoStop}%
\bibitem [{\citenamefont {Alberti}\ \emph {et~al.}(2016)\citenamefont
  {Alberti}, \citenamefont {Robens}, \citenamefont {Alt}, \citenamefont
  {Brakhane}, \citenamefont {Karski}, \citenamefont {Reimann}, \citenamefont
  {Widera},\ and\ \citenamefont {Meschede}}]{Alberti:2016}%
  \BibitemOpen
  \bibfield  {author} {\bibinfo {author} {\bibfnamefont {A.}~\bibnamefont
  {Alberti}}, \bibinfo {author} {\bibfnamefont {C.}~\bibnamefont {Robens}},
  \bibinfo {author} {\bibfnamefont {W.}~\bibnamefont {Alt}}, \bibinfo {author}
  {\bibfnamefont {S.}~\bibnamefont {Brakhane}}, \bibinfo {author}
  {\bibfnamefont {M.}~\bibnamefont {Karski}}, \bibinfo {author} {\bibfnamefont
  {R.}~\bibnamefont {Reimann}}, \bibinfo {author} {\bibfnamefont
  {A.}~\bibnamefont {Widera}},\ and\ \bibinfo {author} {\bibfnamefont
  {D.}~\bibnamefont {Meschede}},\ }\bibfield  {title} {\enquote {\bibinfo
  {title} {{Super-resolution microscopy of single atoms in optical
  lattices}},}\ }\href {https://doi.org/10.1088/1367-2630/18/5/053010}
  {\bibfield  {journal} {\bibinfo  {journal} {New J. Phys.}\ }\textbf {\bibinfo
  {volume} {18}},\ \bibinfo {pages} {053010} (\bibinfo {year}
  {2016})}\BibitemShut {NoStop}%
\bibitem [{\citenamefont {Robens}\ \emph {et~al.}(2017)\citenamefont {Robens},
  \citenamefont {Brakhane}, \citenamefont {Alt}, \citenamefont {Klei{\ss}ler},
  \citenamefont {Meschede}, \citenamefont {Moon}, \citenamefont {Ramola},\ and\
  \citenamefont {Alberti}}]{Robens:2017a}%
  \BibitemOpen
  \bibfield  {author} {\bibinfo {author} {\bibfnamefont {C.}~\bibnamefont
  {Robens}}, \bibinfo {author} {\bibfnamefont {S.}~\bibnamefont {Brakhane}},
  \bibinfo {author} {\bibfnamefont {W.}~\bibnamefont {Alt}}, \bibinfo {author}
  {\bibfnamefont {F.}~\bibnamefont {Klei{\ss}ler}}, \bibinfo {author}
  {\bibfnamefont {D.}~\bibnamefont {Meschede}}, \bibinfo {author}
  {\bibfnamefont {G.}~\bibnamefont {Moon}}, \bibinfo {author} {\bibfnamefont
  {G.}~\bibnamefont {Ramola}},\ and\ \bibinfo {author} {\bibfnamefont
  {A.}~\bibnamefont {Alberti}},\ }\bibfield  {title} {\enquote {\bibinfo
  {title} {{High numerical aperture (NA = 092) objective lens for imaging and
  addressing of cold atoms}},}\ }\href {https://doi.org/10.1364/OL.42.001043}
  {\bibfield  {journal} {\bibinfo  {journal} {Opt. Lett.}\ }\textbf {\bibinfo
  {volume} {42}},\ \bibinfo {pages} {1043} (\bibinfo {year}
  {2017})}\BibitemShut {NoStop}%
\bibitem [{\citenamefont {Kao}\ \emph {et~al.}(2017)\citenamefont {Kao},
  \citenamefont {Tang}, \citenamefont {Burdick},\ and\ \citenamefont
  {Lev}}]{Kao:2017}%
  \BibitemOpen
  \bibfield  {author} {\bibinfo {author} {\bibfnamefont {W.}~\bibnamefont
  {Kao}}, \bibinfo {author} {\bibfnamefont {Y.}~\bibnamefont {Tang}}, \bibinfo
  {author} {\bibfnamefont {N.~Q.}\ \bibnamefont {Burdick}},\ and\ \bibinfo
  {author} {\bibfnamefont {B.~L.}\ \bibnamefont {Lev}},\ }\bibfield  {title}
  {\enquote {\bibinfo {title} {{Anisotropic dependence of tune-out wavelength
  near Dy 741-nm transition}},}\ }\href {https://doi.org/10.1364/OE.25.003411}
  {\bibfield  {journal} {\bibinfo  {journal} {Opt. Express}\ }\textbf {\bibinfo
  {volume} {25}},\ \bibinfo {pages} {3411} (\bibinfo {year}
  {2017})}\BibitemShut {NoStop}%
\bibitem [{\citenamefont {Becher}\ \emph {et~al.}(2018)\citenamefont {Becher},
  \citenamefont {Baier}, \citenamefont {Aikawa}, \citenamefont {Lepers},
  \citenamefont {Wyart}, \citenamefont {Dulieu},\ and\ \citenamefont
  {Ferlaino}}]{Becher:2018}%
  \BibitemOpen
  \bibfield  {author} {\bibinfo {author} {\bibfnamefont {J.~H.}\ \bibnamefont
  {Becher}}, \bibinfo {author} {\bibfnamefont {S.}~\bibnamefont {Baier}},
  \bibinfo {author} {\bibfnamefont {K.}~\bibnamefont {Aikawa}}, \bibinfo
  {author} {\bibfnamefont {M.}~\bibnamefont {Lepers}}, \bibinfo {author}
  {\bibfnamefont {J.~F.}\ \bibnamefont {Wyart}}, \bibinfo {author}
  {\bibfnamefont {O.}~\bibnamefont {Dulieu}},\ and\ \bibinfo {author}
  {\bibfnamefont {F.}~\bibnamefont {Ferlaino}},\ }\bibfield  {title} {\enquote
  {\bibinfo {title} {{Anisotropic polarizability of erbium atoms}},}\ }\href
  {https://doi.org/10.1103/PhysRevA.97.012509} {\bibfield  {journal} {\bibinfo
  {journal} {Phys. Rev. A}\ }\textbf {\bibinfo {volume} {97}},\ \bibinfo
  {pages} {012509} (\bibinfo {year} {2018})}\BibitemShut {NoStop}%
\bibitem [{\citenamefont {Ravensbergen}\ \emph {et~al.}(2018)\citenamefont
  {Ravensbergen}, \citenamefont {Corre}, \citenamefont {Soave}, \citenamefont
  {Kreyer}, \citenamefont {Tzanova}, \citenamefont {Kirilov},\ and\
  \citenamefont {Grimm}}]{Ravensbergen:2018}%
  \BibitemOpen
  \bibfield  {author} {\bibinfo {author} {\bibfnamefont {C.}~\bibnamefont
  {Ravensbergen}}, \bibinfo {author} {\bibfnamefont {V.}~\bibnamefont {Corre}},
  \bibinfo {author} {\bibfnamefont {E.}~\bibnamefont {Soave}}, \bibinfo
  {author} {\bibfnamefont {M.}~\bibnamefont {Kreyer}}, \bibinfo {author}
  {\bibfnamefont {S.}~\bibnamefont {Tzanova}}, \bibinfo {author} {\bibfnamefont
  {E.}~\bibnamefont {Kirilov}},\ and\ \bibinfo {author} {\bibfnamefont
  {R.}~\bibnamefont {Grimm}},\ }\bibfield  {title} {\enquote {\bibinfo {title}
  {{Accurate Determination of the Dynamical Polarizability of Dysprosium}},}\
  }\href {https://doi.org/10.1103/PhysRevLett.120.223001} {\bibfield  {journal}
  {\bibinfo  {journal} {Phys. Rev. Lett.}\ }\textbf {\bibinfo {volume} {120}},\
  \bibinfo {pages} {223001} (\bibinfo {year} {2018})}\BibitemShut {NoStop}%
\end{thebibliography}%
\end{document}